\documentclass[12pt]{article}
\usepackage{latexsym}
\def\beq{\begin{equation}}
\def\eeq{\end{equation}}
\def\bey{\begin{eqnarray}}
\def\eey{\end{eqnarray}}

\def\and{{\rm and\ }}
\def\be{\begin{equation}}
\def\ee{\end{equation}}

\def\spose#1{\hbox to 0pt{#1\hss}}
\def\lta{\mathrel{\spose{\lower 3pt\hbox{$\sim$}}
    \raise 2.0pt\hbox{$<$}}}
\def\gta{\mathrel{\spose{\lower 3pt\hbox{$\sim$}}
    \raise 2.0pt\hbox{$>$}}}

\input epsf

\title{Expectations for the Deep Impact collision from cometary nuclei modelling}

\author{
\\
O. Mousis$^1$\thanks{E-mail: Olivier.Mousis@obs-besancon.fr}, U. Marboeuf$^1$,
J.-M. Petit$^1$ \& J. Klinger$^2$
\\[3mm]
\\
$^1$ Observatoire de Besan\c{c}on, CNRS-UMR 6091, BP 1615,\\
25010 Besan\c{c}on
Cedex, France\\
$^2$ Tizin, 38210 Tullins, France\\
\\
Submitted to MNRAS, June 1st, 2005.\\
Revised version, June 15th, 2005 
\\
}
 \date{\rule{0mm}{0mm}}

\begin{document}
\maketitle

\begin{abstract}
Using the cometary nucleus model developed by Espinasse et al. (1991), we
calculate the thermodynamical evolution of Comet 9P/Tempel 1 over a period of
360 years.
Starting from an initially amorphous cometary nucleus which incorporates an icy
mixture of H$_2$O and CO, we show that, at the time of $Deep$ $Impact$
collision, the crater is expected to form at depths where ice is in its
crystalline form. Hence, the subsurface exposed to space should not be
primordial. We also attempt an order-of-magnitude estimate of the heating and
material ablation effects on the crater activity caused by the 370 Kg
projectile released by the $DI$ spacecraft. We thus show that heating effects
play no role in the evolution of crater activity. We calculate that the CO
production rate from the impacted region should be about 300-400 times
higher from the crater resulting from the impact with a 35 m ablation than over
the unperturbed nucleus in the immediate post-impact period. We also show that
the H$_2$O production rate is decreased by several orders of magnitude at the
crater base just after ablation.
\end{abstract}

\section{Introduction}
On July 4 2005, the 370 Kg projectile released by the dual $Deep$ $Impact$
spacecraft will collide with short period comet 9P/Tempel 1 one day before
perihelion at a velocity of 10.2 km.s$^{-1}$. At the same time, the flyby
spacecraft will take optical images and near-IR spectra of the nucleus, the
crater formation process, the resultant crater and ejecta, and any induced
outgassing. The main scientific goals of this mission are to infer the bulk
properties of the cometary nucleus, better understand its internal structure,
compare outgassing composition before and after impact and derive some
constraints on the crater formation process.

In the present work, we use the cometary nucleus model developed by Espinasse
et al. (1991) to examine the plausible current thermodynamical state of Comet
9P/Tempel 1. Starting with an initially amorphous cometary nucleus, we follow
the evolution of Comet 9P/Tempel 1's stratigraphy over a period covering the
last 360 years of its orbital history\footnote{The orbital evolution of this
comet is poorly constrained at earlier epochs (see Section 2.3).}. This allows
us to infer the nature of the ice, which may be amorphous (therefore pristine)
or crystalline, according to the thermal evolution of the nucleus, and that
should be revealed at the crater base after the impact from the $DI$
projectile. We also give order-of-magnitude estimates of the combined effects of the
material ablation and heating on the crater activity caused by the $DI$
impactor. Ablation consists, at the time of the impact, in the removal of an
outer icy layer of the nucleus whose thickness corresponds approximately to the
crater depth. Impact heat is transfered to the crater subsurface immediately
after ablation, and its propagation occurs in the radial direction of the
cometary nucleus, following the approach described by Orosei et
al. (2001). Once these two processes are taken into account, this enables us to
calculate the production rates of the main volatiles (H$_2$O and CO) that are
expected to outgass from the crater.

In Section 2, a brief description of the nucleus model is presented. We also
give some details concerning the impact modelling and the choice of the nucleus
parameters. In Section 3, we calculate the thermodynamical evolution of Comet
9P/Tempel 1 over the last 360 years of its orbital history. We also estimate in
this Section the effects induced by impact heating and ablation on the crater
activity. Section 4 is devoted to conclusions.

\section{Models}

\subsection{The cometary nucleus model}
The nucleus model employed in this work is the 1D model originally developed by
Espinasse et al. (1991). This model takes a sphere (initially homogenous)
composed of a highly porous predefined mixture of water ice (the dominant
constituent) and CO ice in specified proportions. We consider that the mixture
is initially amorphous because of the dynamical evolution of the comet
described in Section 2.3. CO is trapped in the amorphous water ice, with the
remaining, if any, being condensed in the pores. When heated,
the condensed CO sublimates first, then the trapped CO will be released during
the transition from amorphous to crystalline water ice.  The model is based on
the simultaneous resolution for the whole nucleus of a heat diffusion equation
and an equation of gas diffusion in the pore system. It accounts for heat
transmission, gas diffusion, and sublimation/condensation of volatiles within
the matrix, solving the partial differential equations which describe the
nucleus.
However, note that the model used here has been the subject of several
refinements in recent years (Espinasse et al. 1993; Orosei et al. 1995; Orosei
et al. 1999). In particular, with the incorporation of dust and the use of more
appropriate numerical methods, the recent versions of the model provide an
improved description of the interior evolution of the nucleus. On the other
hand, the present work started a few months ago using the original nucleus
model of Espinasse et al. (1991). In these conditions, the imminency of the
$DI$ event did not allow us to take time to incorporate these latter
improvements in our code. Nevertheless, even in its current version, this model
allows us to examine some interesting clues concerning the thermochemical state
of the Comet 9P/Tempel 1, both before and after the collision.

\subsection{Impact modelling}

In this work, the collision between the $DI$ projectile and Comet 9P/Tempel 1
is characterized in terms of 1) the ablation of cometary material due to the
cratering event and 2) the amount of kinetic energy delivered in the form of
heat to the target by the impactor itself. Ablation consists in the removal, at
the time of the impact, of an outer icy layer of the cometary nucleus, whose
thickness corresponds roughly to the crater depth due to the projectile
impact. Immediately after ablation, impact heat is transfered to the crater
base and its propagation in the cometary nucleus is described following the
approach of Orosei et al. (2001). Because of the 1-D nature of the model,
lateral heat diffusion is neglected. As pointed out by Orosei et al. (2001),
this method has some limitations. In particular, the depth at which the impact
can thermally alter the comet must be considered as an upper limit. In the
present case, the distribution of impact heat results in a homogeneous energy
density deposited in the crater subsurface within a cylinder whose
cross-section and depth correspond respectively to the crater diameter and the
size of the projectile. Once these two physical processes are taken into
account in our model, it is possible to give an order-of-magnitude estimate of
the combined effects of the material ablation and heating on the crater
activity caused by the $DI$ projectile.

\subsection{Choice of parameters}
Table 1 summarizes the set of parameters selected for the nucleus of Comet
9P/Tempel 1. Two different CO:H$_2$O initial ratios were selected for the
composition of the comet, namely 5 \% for the first model (\textit{CO-poor
comet}), and 15 \% for the second model (\textit{CO-rich comet}).
{
In the model, any CO in excess of 10\% appears as condensed CO ice.
}
The resulting
densities, which have been determined using a porosity of 0.7, are compatible
with the estimated density of Comet 9P/Tempel 1 (Belton et al. 2005). The
nucleus radius and geometric albedo are taken from Lamy et al. (2004) and Lisse
et al. (2005a) respectively. The initial pore radius has been fixed to
10$^{-5}$ m, following De Sanctis et al. (1999). Emissivity $\epsilon$ has been
chosen to be close to 1 since the nucleus can be modelled as a black body due
to the high porosity of its material. The initial temperature of 30K for the
whole nucleus corresponds to the equilibrium temperature for a highly
conducting black sphere orbiting at about 90 AU from the Sun.
{
Following Espinasse (1989) and Corradini (1997), we used Russel (1935) formula
to compute the bulk conductivity from those of amorphous and crystaline water
ice (Klinger, 1980).
}

Table 2 describes the orbital history of Comet 9P/Tempel 1 over the last 360
years. The parameters given here are averages of the orbital data from the
numerical simulations reported on the NASA web-site\footnote
{http://deepimpact.jpl.nasa.gov/science/tempel1-orbitalhist.html}. Due to the
limitations of these simulations, the orbital history of Comet 9P/Tempel 1 is
poorly determined at epochs earlier than~$\sim$~1640. Hence, we opted to
perform our calculations of its thermodynamic evolution during the period
covered by Table 2. In this work, following papers published on the origin of
short period comets, we assume that Comet 9P/Tempel 1 originates from the
Kuiper belt region (see Morbidelli (1999) for a review). Since the Kuiper belt
was probably rather cold at the epoch of the Solar nebula formation, this comet
must have formed from pristine ices directly originating from the presolar
cloud, as predicted by Mousis et al. (2000), thus justifying the assumption of
an initially amorphous nucleus.

Since the amount of energy transported by the projectile is quite small
($E_{kinetics}$ = 1.9 $\times$
10$^{10}$ J, with a projectile mass of 370 Kg and velocity of 10.2
km.s$^{-1}$), we deliberately adopted a set of impact parameters that optimize
the influence of collisional heating on the nucleus thermodynamical
behavior.
{
Adopted values of 50 m for the crater size and 35 m for its depth derive from
the cratering scenarios established by Schultz \& Anderson (2005) for the $DI$
collision, and have been chosen to favor a higher energy density and a deeper
energy deposition.
}
We also optimized the energy transfer due to the impact by assuming
that 90 $\%$ of the projectile's kinetic energy is deposited as heat in the
nucleus, with the remaining 10 $\%$ being distributed into the ejected
fragments.

\begin{table}
            \caption[]{Nucleus model initial parameters}
      \begin{center}
         {\setlength{\tabcolsep}{0.1cm}
         \begin{tabular}{lcc}
	    \hline
            \hline
            \noalign{\smallskip}
	      Parameter & CO-poor comet  & CO-rich comet   \\
             \noalign{\smallskip}
             \hline
             \noalign{\smallskip}
	    Structural &  &  \\
	    ~~~Radius (R) & \multicolumn{2}{c}{3.1 km} \\
	    ~~~Bulk density ($\rho$) & 283 Kg m$^{-3}$& 296 Kg m$^{-3}$  \\
	    ~~~Porosity ($p$) & \multicolumn{2}{c}{0.7}  \\
	    ~~~Pore radius ($r_0$) & \multicolumn{2}{c}{10 $\mu$m}  \\
	    Material & & \\
	    ~~~Fraction of trapped CO & 0.05 & 0.15 \\
	    ~~~Albedo (A) & \multicolumn{2}{c}{0.04} \\
	    ~~~Emissitivity ($\epsilon$) & \multicolumn{2}{c}{0.95} \\
	    ~~~Temperature & \multicolumn{2}{c}{30 K} \\
	    
           \noalign{\smallskip}
            \hline
         \end{tabular}}
       \end{center}
      \end{table}

     \begin{table}
            \caption[]{Orbital parameters (semi-major axis a, eccentricity e) for the last 360 years. Epochs refer to the beginning of each orbit.}
      \begin{center}
         {\setlength{\tabcolsep}{0.1cm}
        \begin{tabular}{lccclccc}
	    \hline
           \hline
            \noalign{\smallskip}
	      & a (AU)  & e & Epochs & & a (AU)  & e & Epochs\\
             \noalign{\smallskip}
             \hline
             \noalign{\smallskip}
	    Orbit 1 & 3.92 & 0.27 & 1645 & Orbit 6 & 3.29 & 0.46 & 1870 \\ 
	    Orbit 2 & 3.71 & 0.32 & 1668 & Orbit 7 & 3.48 & 0.4   & 1882 \\
	    Orbit 3 & 3.42 & 0.43 & 1704 & Orbit 8 & 3.24 & 0.47 & 1941 \\
	    Orbit 4 & 3.21 & 0.45 & 1776 & Orbit 9 & 3.12 & 0.52 & 1954\\  
	    Orbit 5 & 3.18 & 0.5 & 1787 & & &\\
           \noalign{\smallskip}
            \hline
         \end{tabular}}
       \end{center}
      \end{table}

\section{Results}

\subsection{Thermodynamical evolution of Comet 9P/Tempel 1}

Figure 1 shows a plausible thermodynamical evolution of the nucleus of a
CO-poor comet over a period covering the last 360 years in the orbit of Comet
9P/Tempel 1. Starting with an initially amorphous nucleus, the
amorphous/crystalline phase transition takes place at the same time as CO
emission starts. This phase transition occurs deeper as soon as the comet is
injected into an orbit closer to the Sun. At the end of the 360 year period,
the surface ablation reaches a depth of 34 m and the lower limit of the
complete crystalline layer is located 427.8 m below the surface. An
intermediary layer containing a mixture of both amorphous and crystalline
H$_2$O ice, and solid CO (trapped in the amorphous matrix or condensed in the
pores), is formed between the crystalline layer and the undifferentiated
core. The upper limit of the totally amorphous layer is located 485 m below
the surface.

Figure 2 is similar to figure 1 but instead examines a CO-rich comet. At the
end of the calculations, the surface ablation reaches a depth of 37.4 m and the
lower limit of the complete crystalline layer is located 201 m below the
surface. The upper limit of the undifferentiated core is located at a depth of
215 m. In that case, the crystalline layer is much thinner than in the
CO-poor comet case. As stated by Espinasse et al. (1991), the presence of a
larger amount of CO in the nucleus tends to slow the propagation of the
amorphous to crystalline phase transition because a large fraction of the
incoming energy is used to vaporize the extra CO. We have performed the same
calculations as in Figs. 1 and 2 but for porosities of 0.3. Adopting a lower
porosity tends to increase the amplitude of crystallization. Indeed, the lower
limit of the crystalline layer is found to be deeper (555.9 m for the CO-poor
comet and 236.2 m for the CO-rich comet), even if, in both cases, the value of
the surface ablation depth after 360 years of orbital evolution is lower (12.9
m for the CO-poor comet and 14.2 m for the CO-rich comet).
\begin{figure}
\centering
\vspace{85mm}
\includegraphics{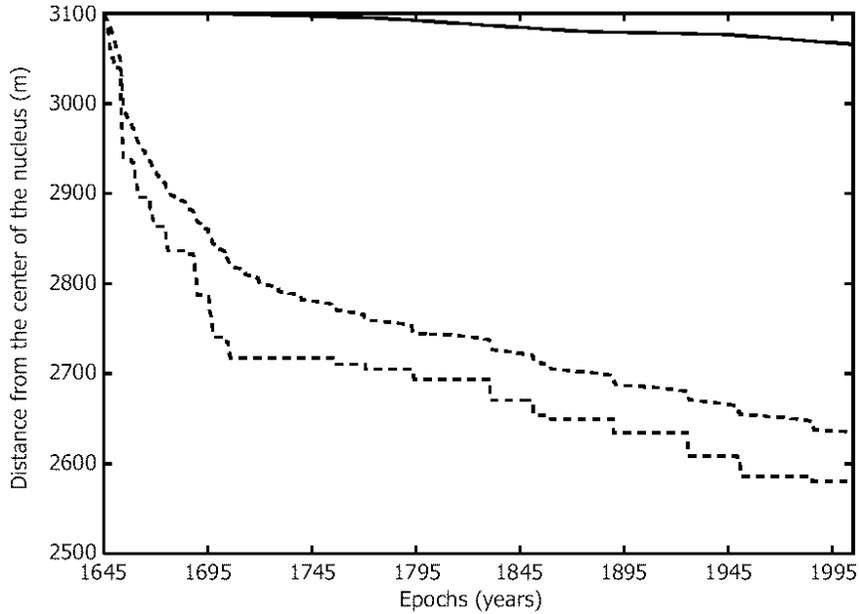}
\caption{Evolution of the CO-poor comet's stratigraphy over the last 360 years,
assuming an initially amorphous nucleus. The lines represent, from top to
bottom, the surface, the lower limit of complete crystalline layer and the
upper limit of the totally amorphous ice layer as a function of orbital
evolution. The lower limit of total depletion of trapped CO is the same as the
lower limit of the complete crystalline layer.}
\end{figure}

\begin{figure}
\centering
\vspace{85mm}
\includegraphics{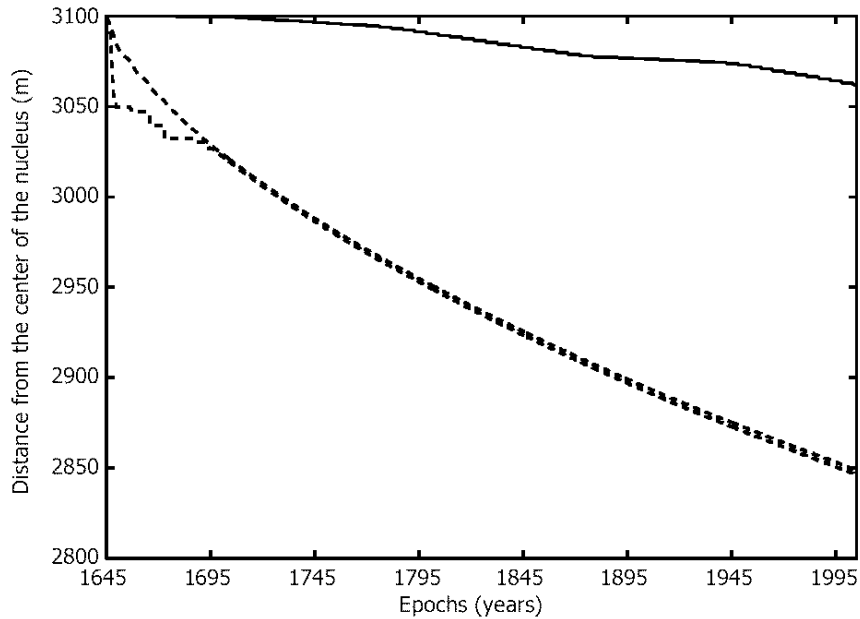}
\caption{Same as in Fig. 1 but for the case of the CO-rich comet.}
\end{figure}

The calculations presented here suggest that, even if the orbital history of
Comet 9P/Tempel 1 is not taken into account at epochs prior to the last 360
years, the thickness of the crystalline layer exceeds greatly the depth of the
crater expected to be formed by the $DI$ projectile. This suggests that the
nucleus material revealed at the crater base should not be primordial. A phase
transition between amorphous and crystalline H$_2$O ice has occurred over a
minimum of several hundred of metres depth from the surface of the nucleus
during the thermal history of the comet and the abundance of CO present in this
layer no longer reflects its primordial value at the time of comet formation.

Figure 3 shows the H$_2$O and CO production rates (Q(H$_2$O) and Q(CO)
respectively) for both CO-rich and CO-poor comets. It can be seen that the
production rates at perihelion vary as a function of the orbital history of
Comet 9P/Tempel 1.
{
Production rates at the beginning of the simulation are meaningless, as they
result from the abrupt injection from a distant orbit into an orbit close to
the Sun. After about 3 orbits, thermal equilibrium is reached and production
rates are informative.
}
The H$_2$O production rates of both CO-rich and CO-poor comets are practically
identical. On its current orbit, the 
H$_2$O production rate given by the two models is $\sim$ 1.5~$\times$~10$^{-3}$
mole.m$^{-2}$.s$^{-1}$ at perihelion,
i.e.
$\sim$~1.1~$\times$~10$^{29}$ molecules.s$^{-1}$ for the whole nucleus.
This latter value is around 5 times higher than the gas production
rate derived from measurements of the activity of Comet 9P/Tempel 1 made by
A'Hearn et al. (1995).
Such a difference can easily be explained by considering that the activity is
concentrated on the subsolar point (Enzian et al.  1999). Additionally, dust 
accumulation on the surface may also modulate the activity (De Santis et
al. 1999). This latter possibility is confirmed by the recent work of
Lisse~et~al.~(2005b) who derived, using a thermal model, that about 9 $\pm$ 2
$\%$ of the surface of Comet 9P/Tempel 1 is active at perihelion.

As shown in Fig. 3, the evolution of the CO production rate varies as a
function of the initial CO:H$_2$O ratio adopted in the nucleus. At the very
beginning, the peaks in the production rates are more important in the CO-rich
comet than in the CO-poor comet. With time, the increasing thickness of the
crystalline layer tends to lower Q(CO) of CO-poor comet. The quasi-constant CO
production rate in the case of the CO-rich comet results from the regular
sublimation of the portion of CO condensed in the pores. For the CO-poor comet,
peaks in Q(CO) are strongly correlated with peaks of crystalization (steep
slope in the lower limit of crystalline zone in Fig. 1), which in turn occur
slightly after the orbit has changed for a new perihelion distance. At the time
of the $DI$ encounter, which is about one day before perihelion, Q(CO) is about
5~$\times$~10$^{-6}$ mole.m$^{-2}$.s$^{-1}$ and
2~$\times$~10$^{-5}$~mole.m$^{-2}$.s$^{-1}$ in the cases of CO-poor and CO-rich
comets, respectively.

{
The most important caveat is that a strong surface ablation before impact
(WHY?) could drastically decrease the depth of crystalline ice.
In addition, our model does not consider the latidudinal variation of the solar
insolation.
}

\begin{figure}
\centering
\vspace{85mm}
\includegraphics{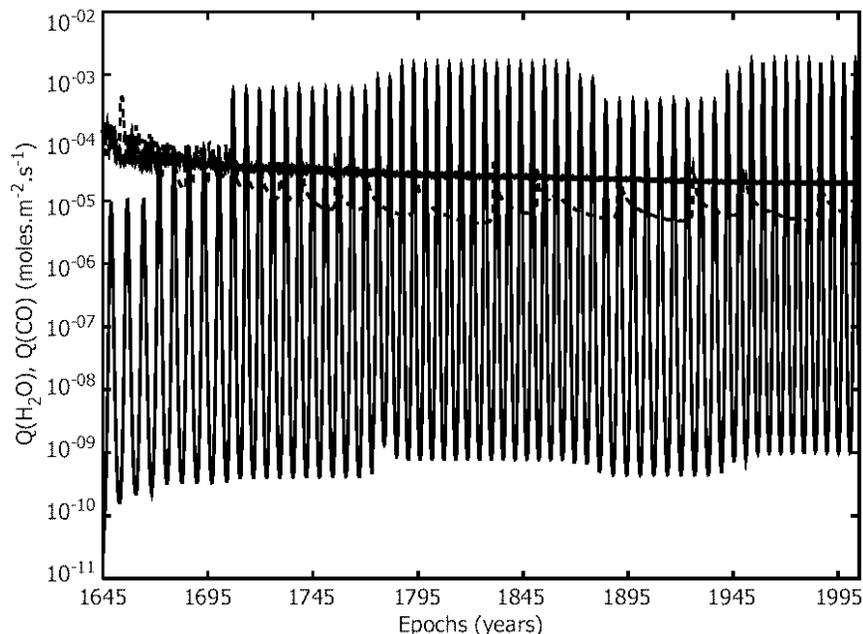}
\caption{Production rates over the last 360 years. Graphs represent Q(H$_2$O)
for both models (solid line), Q(CO) for CO-rich comet (bold solid line) and for
CO-poor comet (dashed line).}
\end{figure}

\subsection{Effects of impact heating and ablation}

We calculated the heating induced by the energy transfer to the crater base,
once ablation was over. Adopting the impact parameters given in Section 2.3
which favor a higher energy density, this results in a slight temperature
elevation in the crater subsurface ($\sim$ a few K), compared to the nucleus
temperature at the same depth just before the collision
{
($\sim$ 140 K)
}.
As our nucleus model shows, this temperature increase is too low to ease the
subsequent sublimation of the exposed H$_2$O ice at the crater base. Hence,
heating effects due to the impact play no role in the evolution of the crater
activity.

However, ablation of cometary material due to the cratering leads to several
important consequences.
Just after ablation, and due to the delay required for the subsurface to reach
an equilibrium state with insolation, the freshly exposed ice remains at a
lower temperature than the average at the surface for a long time.
As a result, the H$_2$O production rate Q(H$_2$O) is decreased by several
orders of magnitudes in the crater zone just after ablation
{
(Q(H$_2$O) is about 10$^{-5}$ and 10$^{-6}$ mole.m$^{-2}$.s$^{-1}$ for CO-rich
  and CO-poor comets respectively, with a 35 m ablation depth)
}.
With time, Q(H$_2$O) of the crater increases, such that approximately between 12 and 24 hours after the impact, it is
still half that calculated for both CO-rich and CO-poor unperturbed comets.

Figure 4 describes the evolution of CO production rates Q(CO) for both CO-rich
and CO-poor comets following the impact, for two different ablation depths,
namely 35 m (our nominal case) and 10 m. These production rates reflect the
release through the crater surface of gaseous CO (mainly present in the pores
of the nucleus' crystalline layer). CO production rates for the unperturbed
CO-rich and CO-poor comets are also given for comparison. It can be seen that,
just after a 35 m ablation due to the impact, Q(CO) at the crater base is
enhanced by a factor of about 300-400 times the value for the unperturbed
nucleus in both models (1.4~$\times$~10$^{-3}$ mole.m$^{-2}$.s$^{-1}$ and
8.4~$\times$~10$^{-3}$~mole.m$^{-2}$.s$^{-1}$ for Q(CO) of CO-poor and CO-rich
comets, respectively).
If a 10 m ablation is considered, Q(CO) at the crater base of both models is
enhanced by a factor of $\sim$ 100 times the value for the unperturbed nucleus.
{
Values of the production rates strongly depend on the porosity of the nucleus.
For a porosity of 0.3, Q(CO) is increased by about one order
of magnitude both before and after the impact.
}
Starting with a much higher rate than Q(H$_2$O) at the crater base just after
its formation, the value of Q(CO) decreases with time and meets that of
Q(H$_2$O) about 1 hour after the impact for a 35 m ablation.
Moreover, Q(CO) is still
twice that of
the unperturbed comet $\sim$ 48 hours after the impact.

\begin{figure}
\centering
\vspace{85mm}
\includegraphics{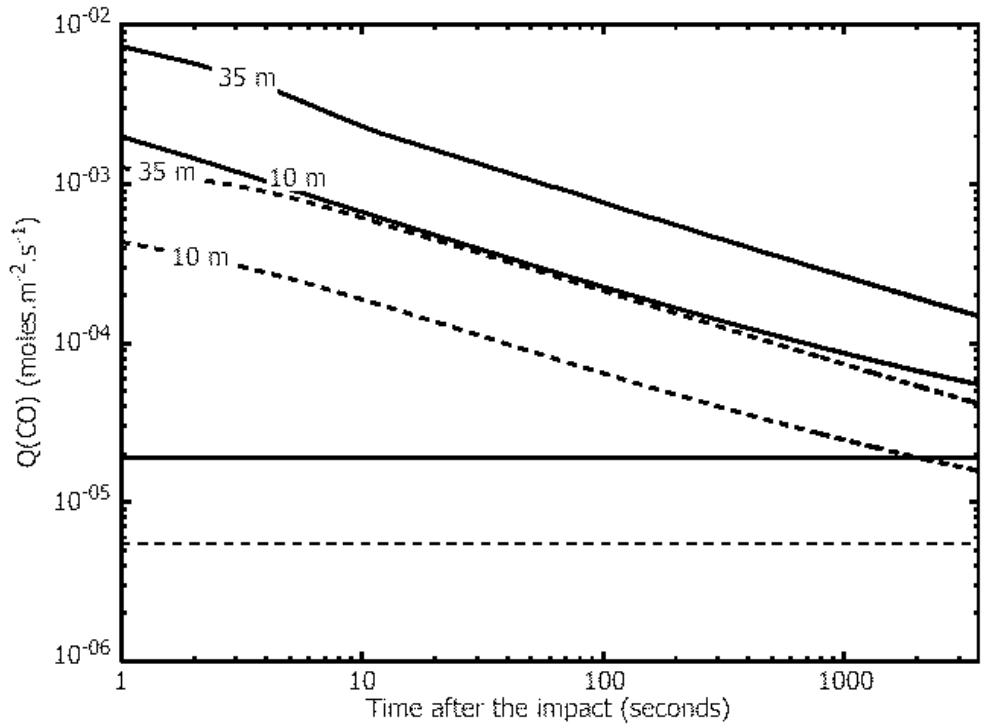}
\caption{Evolution of Q(CO) for one hour after the impact. Horizontal solid and
dashed lines correspond to Q(CO) of the unperturbed CO-rich and CO-poor comets,
respectively. From top to bottom, the oblique solid lines represent the
evolution of Q(CO) of CO-rich comet after the impact with ablations of 35 m and
10 m, respectively. From top to bottom, the oblique dashed lines represent the
evolution of Q(CO) of the CO-poor comet after the impact with ablations of 35 m
and 10 m, respectively.}
\end{figure}

\section{Conclusions}

Using the 1D cometary nucleus model originally developed by Espinasse et
al. (1991), we have elaborated two different models, a CO-poor comet and a
CO-rich comet, which could resemble to Comet 9P/Tempel 1. We calculated the
thermodynamical evolution of these two models over the last 360 years, a period
during which the orbital parameters of Comet 9P/Tempel 1 are fairly well
determined. Starting with an initially amorphous nucleus, we show that, at the
time of impact, the thickness of the crystalline layer greatly exceeds the
depth of the crater expected to be formed by the $DI$ projectile. Hence, the
fresh ice newly exposed at the crater base should not be primordial since it
has suffered a phase transition and a composition change during the thermal
history of the comet.

We also calculated the combined effects of the material ablation and heating on
the crater activity caused by the $DI$ projectile. Ablation consisted here in
the removal of an outer icy layer of the cometary nucleus at the time of the
impact. Impact heat is transfered to the crater subsurface immediately after
ablation as a homogeneous energy density deposited within a cylinder whose
cross-section and depth correspond to the crater diameter and the size of the
projectile, respectively. As a result, due to the very limited temperature
elevation ($\sim$ a few K) of the crater subsurface, we show that impact
heating play no role in the evolution of the crater activity. On the other
hand, we calculate that the H$_2$O production rate is decreased by several
orders of magnitude just after a 35 m ablation
{
(Q(H$_2$O) is about 10$^{-5}$ and 10$^{-6}$ mole.m$^{-2}$.s$^{-1}$ for CO-rich
  and CO-poor comets, respectively)
}.
For the same ablation depth, we also calculate that the CO production rate is
enhanced by a factor of about 300-400 times the value for the unperturbed
nucleus in both models in the immediate post-impact period.

Since the major effect of the impact is the ablation of the top layers,
allowing a huge increase of CO production rate, these results are not sensitive
to the exact impact parameters we have used. In particular, an important
decrease of deposited energy, potentially due to a lower energy transfer
efficiency would not affect our conclusions.

We are aware of the limitations of our model, this latter being based on the
"fast-rotator" approximation. However, it has been shown that it is a good
assumption for the interior of the nucleus, and also for modeling the nucleus
activity close to the Sun (Prialnik 2002). The incorporation of dust in our
calculations would have also provided a more realistic description of the
thermodynamical evolution of the nucleus.
{
In particular, the dust trapped just beneath the nucleus surface would form,
eventually, a dust mantle, which, in turn, would affect the rate of heat and
gas flow at the surface (Prialnik 2002). This would also modify the local pore
size distribution.
}
On the other hand, the production rates calculated here are orders-of-magnitude
estimates and predictions for the evolution of these quantities in the
post-impact period are relative values. Moreover, since the thermal
conductivity of dust is higher than that of amorphous ice, its presence in the
nucleus would rather ease the amorphous to crystalline phase transition, thus
leading to a thicker crystalline layer at the term of our calculations of the
360 years of orbital evolution.

\section*{Acknowledgments}
{
We thank Sylvie Espinasse who made helpful suggestions about the use of her
code, Jonathan Horner for valuable comments on the manuscript and Anny-Chantal
Levasseur-Regourd for profitable discussions.
}

\end{document}